\documentclass[12pt]{iopart}

\usepackage{url}
\usepackage[]{units}
\usepackage{graphicx}
\usepackage{epstopdf}
\begin{document}
\title[]{Globally coherent short duration magnetic field transients and their effect on ground based gravitational-wave detectors}

\author{Izabela~Kowalska-Leszczynska$^{1}$,
Marie-Anne~Bizouard$^{2}$,
Tomasz Bulik$^{1}$,
Nelson~Christensen$^{3,4}$,
Michael~Coughlin$^{5}$,
Mark Go{\l}kowski$^{6}$,
Jerzy Kubisz$^{7}$,
Andrzej Kulak$^{8}$,
Janusz Mlynarczyk$^{8}$,
Florent Robinet$^{2}$,
Max Rohde$^{3}$}
\address{$^{1}$Astronomical Observatory Warsaw University, 00-478 Warsaw, Poland}
\address{$^{2}$LAL, Universit\'e Paris-Sud, CNRS/IN2P3, Universit\'e Paris-Saclay, 91400 Orsay, France \\}
\address{$^{3}$Physics and Astronomy, Carleton College, Northfield, MN 55057, USA}
\address {$^{4}$Universit\'e C\^ote d'Azur, Observatoire C\^ote d'Azur, CNRS, ARTEMIS UMR 7250, 06304 Nice, France}
\address{$^{5}$Department of Physics, Harvard University, Cambridge, MA 02138, USA}
\address{$^{6}$Department of Electrical Engineering, University of Colorado Denver, Denver, CO 80204, USA}
\address{$^{7}$Astronomical Observatory, Jagiellonian University, Krakow, Poland}
\address{$^{8}$AGH University of Science and Technology, Department of Electronics, Krakow, Poland}

\begin{abstract}
It has been recognized that the magnetic fields from the Schumann resonances could affect the search for a stochastic gravitational-wave background by LIGO and Virgo. Presented here are the observations of short duration magnetic field transients that are coincident in the magnetometers at the LIGO and Virgo sites. Data from low-noise magnetometers in Poland and Colorado, USA, are also used and show short duration magnetic transients of global extent. We measure at least 2.3 coincident (between Poland and Colorado) magnetic transient events per day where one of the pulses exceeds 200 pT. Given the recently measured values of the magnetic coupling to differential arm motion for Advanced LIGO, there would be a few events per day that would appear simultaneously at the gravitational-wave detector sites and could move the test masses of order \unit[$10^{-18}$]{m}. We confirm that in the advanced detector era short duration transient gravitational-wave searches must account for correlated magnetic field noise in the global detector network.
\end{abstract}

%
%

\section{Introduction}
The biggest and most sensitive gravitational-wave observatories in the world are presently the two twin interferometers placed in the United States, Advanced LIGO~\cite{0264-9381-32-7-074001,0264-9381-27-8-084006} and the European Virgo~\cite{0264-9381-32-2-024001}, located in Italy. The coming years will also see the arrival of the Japanese gravitational-wave detector, KAGRA~\cite{0264-9381-29-12-124007}, as well as a third LIGO detector that is to be located in India~\cite{doi:10.1142/S0218271813410101}. The Advanced LIGO detectors recently observed gravitational waves from the coalescences of binary black holes~\cite{PhysRevLett.116.061102,PhysRevLett.116.241103,PhysRevX.6.041015}.

To be able to measure the subtle effects that are caused by gravitational waves, the detectors have to be extremely sensitive instruments. Therefore, they are
also sensitive to many sources of environmental noise that need to be identified in the data and removed by properly defined vetoes~\cite{0264-9381-27-19-194010,0264-9381-32-11-115012,0264-9381-29-15-155002,0264-9381-33-13-134001}.
In order to verify that environmental noise is not corrupting the data, all of the detectors and their local environments are constantly monitored by various instruments such as seismometers, accelerometers, microphones, magnetometers, etc.
In this paper we are focusing on magnetic noise transients that are caused by electromagnetic discharges in our atmosphere. The so-called Schumann resonances 
are characteristic structures in the Earth's electromagnetic spectrum. 
The extremely low frequency (ELF) electromagnetic waves propagating around the Earth can create magnetic fields that are coherent over global distances.
The long-term correlation of magnetic noise between the two LIGO detectors and Virgo was studied previously~\cite{2013PhRvD..87l3009T,PhysRevD.90.023013}.
These studies were done in the context of searching for a stochastic gravitational-wave background (SGWB), where the signal search method (looking for a correlated signal in two detectors) assumes that the noise in the two detectors is non-correlated. 
A search for a SGWB is a major research activity for LIGO and Virgo~\cite{PhysRevLett.113.231101}.
In a search for a SGWB the data from two detectors are correlated over a timescale of many months to year~\cite{PhysRevD.46.5250}; hence the long-term presence of correlated magnetic field noise from the Schumann resonances would create a systematic error in the SGWB signal search.
It has been demonstrated that a search for a SGWB by Advanced LIGO and Advanced Virgo could be corrupted by correlated magnetic field noise~\cite{2013PhRvD..87l3009T,PhysRevD.90.023013}.

Whereas the search for a SGWB is done over very long time periods, searches for transient gravitational-wave events examine timescales from milli-seconds to a few minutes~\cite{PhysRevD.87.022002,PhysRevD.85.122007,PhysRevD.93.042005}. Long compact binary coalescence signals are therefore also included.
There is a similar correlated noise source that should be considered when searches are conducted for short duration transient gravitational waves. Specifically, very powerful atmospheric discharges can produce significant electromagnetic waves that can propagate around the world. These electromagnetic events can be observed as individual short-duration coincident signals rather than the quasi-constant background of the Schumann resonances.

A major motivation for the study presented in this paper was an event that occurred on December 12, 2009 near Corsica \cite{JGRD:JGRD16613}. This powerful discharge, called a super jet, was seen in many low-noise magnetometers around the world. It was also detected by the magnetometers at Virgo (Cascina, Pisa, Italy), LIGO-Hanford (Hanford, Washington, USA), and LIGO-Livingston (Livingston, Louisiana, USA). If it had been seen in the detectors' gravitational channels of two interferometers, then it could have been considered a coincident trigger, mimicking gravitational wave. As we will show below, it is very important to perform a regular search for coincident electromagnetic events and distinguish them from potential real gravitational-wave signals.
In Section \ref{sec:schumann_data} we will explain the nature of Schumann resonances and observational methods that are used to study them. We will introduce and describe data collected with low-noise magnetometers located in Poland and Colorado. Then, in Section \ref{sec:gw_data} we will present gravitational-wave detectors LIGO and Virgo.
We will also use magnetometer data from the LIGO and Virgo sites during the 2009 - 2010 period, when these gravitational-wave detectors were operating in their initial configurations. In Section \ref{sec:analysis} we will present the 
method and tools that were used in our coincident transient magnetic field search followed by the obtained results in Section \ref{sec:results}. Concluding remarks are presented in Section~\ref{sec:conclusions}.
\section{Characteristic of natural ELF fields on earth}
\label{sec:schumann_data}
A relatively high level of natural ELF electromagnetic fields is associated with long-range propagation of electromagnetic waves in the \unit[3--3000]{Hz} frequency range. ELF waves propagate in the waveguide formed by the conductive surface of Earth and the lower ionospheric layers at the altitude of about \unit[75]{km}.  Since the reflective layer is located at the altitude lower than half the wavelength, the Earth-ionosphere waveguide behaves like a low-loss transmission line. The attenuation rate in the lower part of the ELF range is particularly small. At \unit[10]{Hz} it is only \unit[0.25]{dB}/\unit[1000]{km}~\cite{6353166}. It increases with approximately the square root of frequency, reaching a maximum close to the cutoff frequency, which is $\approx$ \unit[1500]{Hz}. The phase velocity, similar to the attenuation, has a dispersive nature; it is equal to about \unit[0.75]{c} at \unit[10]{Hz} and gradually increases with frequency.

The main source of ELF waves in the Earth-ionosphere waveguide are negative cloud-to-ground lightning discharges. They occur as a result of the accumulation of negative charges in the bottom part of a storm cell, leading to the electric breakdown between the cloud and the ground. A single negative cloud-to-ground strike is a short current impulse that lasts for about \unit[75]{$\mu$s} and is associated with the charge flow of about \unit[2.5]{C} in the plasma channel that has the length of \unit[2 to 3]{km}. A typical charge moment change of negative cloud-to-grounds is about \unit[6]{Ckm}~\cite{rakov2003lightning}. It generates an electromagnetic field pulse which spectrum is practically flat up to the cutoff frequency. 

On Earth, mostly in the tropics, about 1000 storm cells are constantly active, and they generate about 50 negative cloud-to-ground  discharges per second. The vertical lightning discharge radiates electromagnetic waves in all directions. They propagate around the world and interfere with each other. As a result, the spectrum of atmospheric noise has a resonant character. This was predicted for the first time by W.O. Schumann~\cite{Sch1951} in 1952. Schumann solved the field equations in the spherical Earth-ionosphere cavity built of the perfectly conducting ground and ionosphere, and obtained the following eigenfrequencies: \unit[10.6, 18.4, 26.0, ...]{Hz}. Due to the dispersive attenuation of the Earth-ionosphere waveguide, the measured resonance frequencies proved to be lower: \unit[8, 14, 20 ...]{Hz} (first measurement in 1960~\cite{balser1960observations}). Because of relatively small quality factors of the Earth-ionosphere cavity, which are equal to about \unit[4, 5, 6, ...]{}, respectively, the Schumann resonance peaks are relatively wide.

Since the cavity is excited by a random Poissonian distribution of negative cloud-to-ground discharges in the storm centers, the field in the Earth-ionosphere waveguide has the character of a Gaussian process. Its coherence time does not exceed one second. The spectral density of the first Schumann resonance peak is about \unit[1]{pT/$\sqrt{Hz}$} and at any given moment it is different at each observation location, due to its different position relative to the world storm centers. The background Schumann resonance field sets the lower limit for the amplitude level which can be detected for other sources of magnetic field background.

Other sources include the positive cloud-to-ground atmospheric discharges. They occur because of positive charge accumulation in the upper parts of extended mesoscale convective systems, which leads to a breakdown. As a result, a significant charge (\unit[$\approx$20]{C}) is removed through the plasma channel that has a length of about \unit[12]{km}. Due to a significant charge moment, which is of order of \unit[250]{Ckm}, these discharges generate strong ELF field pulses, visible above the Schumann resonance background noise, even at large distances. The strongest of them have the charge moments of order of \unit[1000]{Ckm} and generate signals classified as Q-bursts.

These electromagnetic events have characteristic waveforms in which multiple round-the-world propagation is visible. This will be seen in some of the examples to follow in this paper. An initial very rapid pulse can be associated with the wave propagating directly from the source. A wider (in time) subsequent signal can be associated with a round-the-world wave. Since the first impulse is short in time and has a large amplitude, it will be observed with a relatively large signal-to-noise ratio (SNR) and a wide frequency bandwidth. After propagating around the world, the higher frequency components are more attenuated than the lower frequency components, so for the subsequent signals only the low frequency components are clearly visible, especially at the Schumann resonance frequencies.

Very strong discharges also occur between the clouds and the ionosphere. The most common type of such a discharge is associated with a transient luminous event, known as a Sprite. The occurrence of a Sprite above a mesoscale convective system is usually preceded by a positive cloud-to-ground discharge. Sprites are associated with charge flow from the ionosphere to the cloud, which lasts up to a few hundred of milliseconds.  Charge moment change of such discharges can reach several thousand \unit[]{Ckm}~\cite{JGRA:JGRA51647}. Among the most powerful discharges associated with transient luminous events are Gigantic Jets. They usually last shorter than Sprites and can have the charge moment of up to several thousand \unit[]{Ckm}. The rise time of Gigantic Jets is short, so they generate ELF field pulses with very large amplitudes, clearly visible anywhere on Earth. Because of that, the signals associated with them can potentially have an influence on gravitational-wave detectors. However, there are relatively rare, a few to a dozen per year. In Europe, only one case was recorded so far, near Corsica on December 12 2009~\cite{RDS:RDS5799}. The analysis of the impact of the European Gigantic Jet on the Virgo detector is presented as an example in this paper.

\subsection{Observational methods}
\label{sec:obs_methods}
Observations of natural ELF field pulses are carried out using broadband low noise magnetometers. The required sensitivity can be obtained by magnetometers equipped with ferromagnetic antennas based on Faraday effect connected to low noise semiconductor amplifiers. The use of SQUIDs has been limited to occasional technical trials due to technical inconveniences. The sensor parameters that determine the quality of observations are the lower cut-off frequency, the upper cut-off frequency and the noise level. The lower cut-off frequency should be \unit[30]{mHz} or lower in order to  preserve non-distorted waveforms of longer pulses, produced by upper discharges (cloud-to-ionosphere discharges). The upper cut-off frequency (upper frequency limit) determines the time  resolution of a magnetometer. In the case of observations of short impulses, which have broadband spectra and are produced by cloud-to-ground discharges, the upper frequency limit has a direct impact on the amplitude of the recorded impulses. It has small influence on amplitudes of the slow varying fields generated by upper discharges.

The Krakow ELF research team runs the Hylaty ELF station located in a sparsely populated area of Poland, near the Bieszczady National Park~\cite{kulak2014extremely}. The Hylaty observatory is practically free from man-made noise and provides high quality data. Since 2005, we have carried out continuous recordings of two horizontal magnetic field components using the ELA7 magnetometer, which has a frequency range of \unit[30]{mHz} to \unit[60]{Hz}. In 2013 we installed broadband ELA10 magnetometers that have a frequency range of 30 mHz to 300 Hz and are intended for research of ELF field pulses and Schumann Resonances. The new system is characterized by low noise, \unit[0.001]{pT/$\sqrt{Hz}$} at \unit[100]{Hz}, and the timing accuracy is better than \unit[1]{ms}. See \cite{kulak2014extremely} for explicit details on the magnetometers. The same magnetometer model is used at the Hugo station (Hugo State Wildlife Area) in Colorado, USA, installed in 2015, and in southern Patagonia, Argentina, installed in 2016. Due to spectral purity requirements, and to avoid pollution from power sources, the stations are battery powered.

The Hylaty, Hugo and Patagonia stations are located on three continents forming the WERA system (World ELF Radiolocation Array) that enable observation of very strong atmospheric discharges occurring anywhere on Earth~\cite{WERA}. Using inverse solutions developed for this purpose~\cite{JGRD:JGRD16034,RDS:RDS5799,JGRA:JGRA51647} we can determine the current and charge moments of these discharges.
Each of these stations have two ferrite core active
magnetic field antennas, with one oriented to observe magnetic
fields along the North-South direction, and the other oriented
to observe magnetic fields along the East-West direction. 
These instruments are sensitive to the Schumann resonances as well as transient signals from individual lightning discharges. Large peak current discharges are often associated with transient luminous events that occur at stratospheric and mesospheric altitudes~\cite{pasko2010recent} and their transient electromagnetic signals can be observed worldwide. The magnetometers are also sensitive to moderate peak current discharges depending on the relative distance from the receiver and the presence of very long continuing currents which preferentially excite radiation in the ELF band~\cite{GRL:GRL28741}. They have a lower cut-off frequency of \unit[0.03]{Hz} with the overall shape of the spectrum
dominated by $1/f$ noise.

\subsection{Characteristics of the data from the Hylaty and Colorado stations}
\label{sec:characteristics}
The recorded magnetic signals from the Hylaty and Hugo stations can be used for investigations of the influence of natural ELF electromagnetic fields on gravitational-wave detectors. The Hylaty station is located about \unit[1100]{km} from the Virgo detector. The Hugo station is conveniently located midway between the Hanford and Livingston gravitational detectors.

\begin{figure}
  \begin{center}
    \includegraphics[scale=0.6]{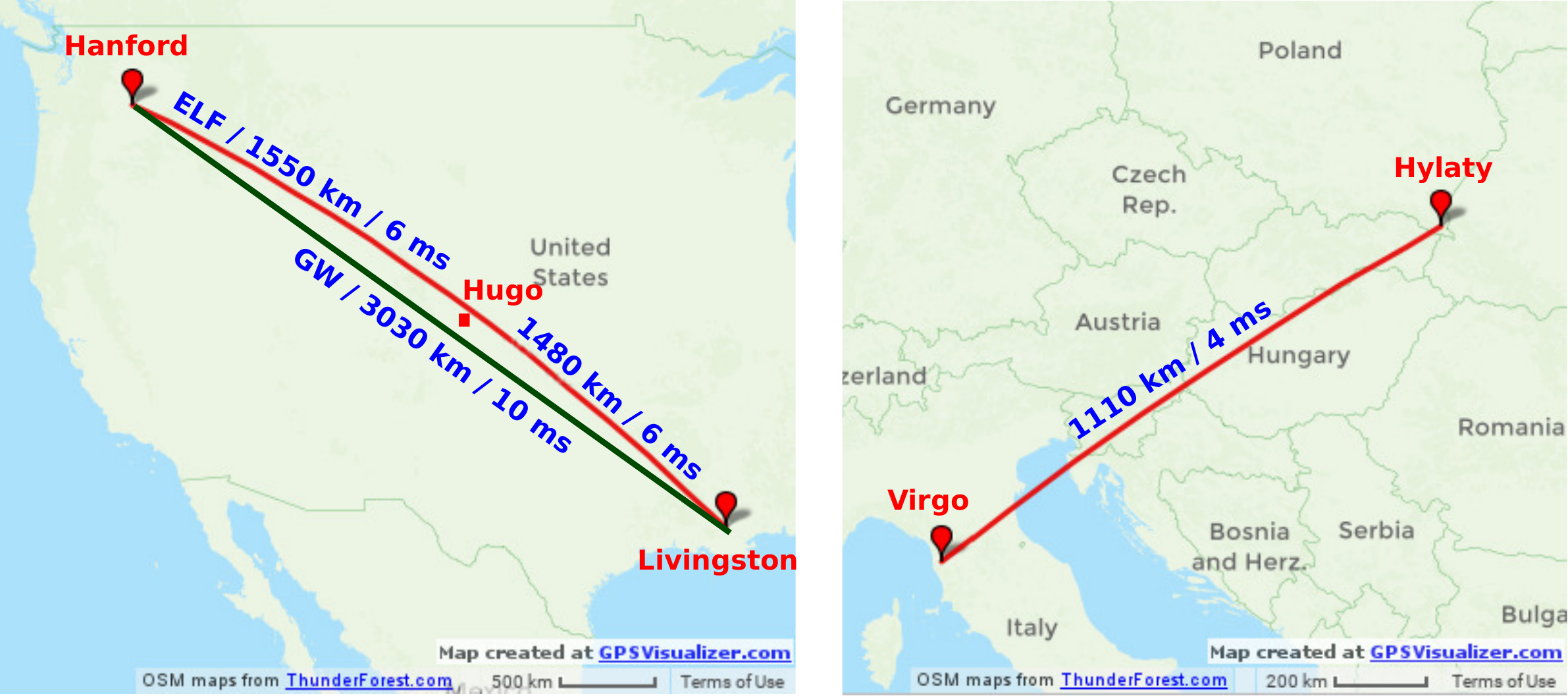}
  \end{center}
  \caption{Left: The location of the Hugo magnetometers and the LIGO-Hanford and LIGO-Livingston gravitational-wave detectors, along with the distance separation and the time of propagation for ELF waves. Right: The location of the Hylaty magnetometers and the Virgo gravitational-wave detector, along with the distance separation and the time of propagation for ELF waves.}
  \label{fig:Hugo}
\end{figure}

Due to the dispersive properties of the Earth-ionosphere waveguide, the velocity of propagation of ELF waves depends on frequency. The observed group velocity for short ELF pulses depends on the upper frequency limit of the magnetometers and for the ELA10 it is about \unit[0.88]{c}. This means that every \unit[1000]{km} introduces a delay of about \unit[3.8]{ms}. It is important to remember that gravitational waves travel in straight lines at the speed of light, straight through the Earth; the ELF signals travel along a great circle. The location of the Hylaty and Hugo magnetometers are displayed in Fig.~\ref{fig:Hugo}, as well as the LIGO and Virgo gravitational-wave detectors. The moment of registration of the signal by a magnetometer is additionally delayed by the group delay of the impulse inside the magnetometer circuitry. The group delay is inversely proportional to the energy bandwidth of a magnetometer~\cite{JGRD:JGRD18081}. In the case of the ELA10 system the group delay of antenna and receiver is \unit[4.2]{ms}. Taking into account the locations of the Hylaty and Hugo stations the maximum difference in time of arrival of impulses does not exceed \unit[30]{ms}. An accurate assessment of the group delay on the output of magnetometers and gravitational detectors is relevant only when investigating the response of the detectors to individual ELF field events, such as Q-bursts and Gigantic Jets.

\section{LIGO/VIRGO data}
\label{sec:gw_data}
The magnetometer data from the LIGO and Virgo sites that are analyzed in this present study came from times during initial LIGO's S6 science run, along with Virgo's VSR2 science run.
LIGO's S6 run spanned the period from July 2009 to October 2010. Virgo operated from July 2009 to January 2010, and then from July 2010 to October 2010. During the initial LIGO and Virgo observations no gravitational waves were detected, but important upper limits on the strength of gravitational-wave backgrounds have been set~\cite{PhysRevLett.113.231101,PhysRevD.85.082002,PhysRevD.87.022002,PhysRevD.89.122003,PhysRevD.85.122007,0004-637X-785-2-119}. While the spectacular detection of gravitational waves had to await the 2015 start of observation by Advanced LIGO~\cite{PhysRevLett.116.061102}, initial LIGO and Virgo also benefited from a wealth of information from environmental sensors~\cite{0264-9381-27-19-194010,0264-9381-32-11-115012,0264-9381-29-15-155002,0264-9381-32-3-035017}. This data from environmental monitors plays an important role for the future operations of Advanced LIGO~\cite{0264-9381-27-8-084006,0264-9381-32-7-074001}, Advanced Virgo~\cite{0264-9381-32-2-024001}, and KAGRA~\cite{0264-9381-29-12-124007}.

Magnetometers are placed at strategic locations around the LIGO and Virgo observatory sites, typically near the gravitational-wave detector test masses. At each location magnetic field measurements are made in the three Cartesian directions. The $x$ and $y$ directions are defined by the interferometer arms, while the $z$ direction is normal to the Earth's surface. Under normal conditions the magnetic fields from the Schumann resonances will only have components parallel to the Earth's surface. The initial LIGO and Virgo magnetometers were mounted inside buildings; the presence of much metal will distort the field direction and allow for observations on the Schumann magnetic fields in the $z$ directed magnetometers as well. 
For the observation of the December 12, 2009 Gigantic Jet the magnetometer used at the Virgo site was a “Broadband Induction Coil Magnetometer”, model MFS-06 by Metronix. For the observation of the Gigantic Jet the LIGO-Livingston magnetometer was a Bartington Mag-03, while the LIGO-Hanford observation was done with a custom made coil magnetometer.

\begin{figure}
  \begin{center}
    \begin{tabular}{cc}
      \includegraphics[scale=0.3]{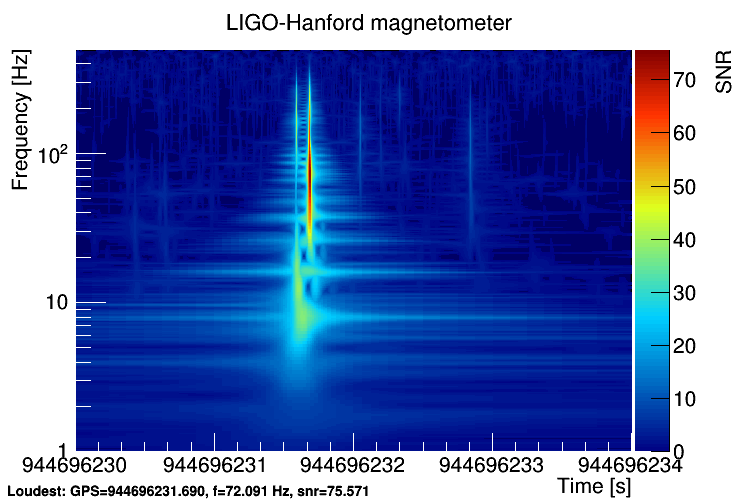}&
      \includegraphics[scale=0.3]{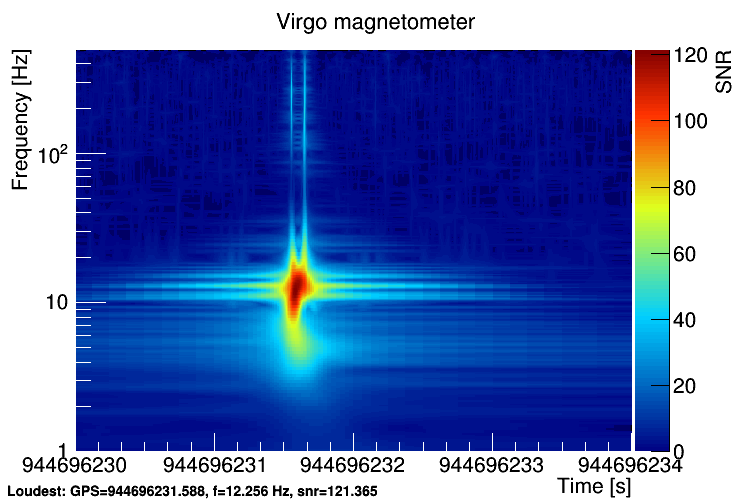} \\
      \includegraphics[scale=0.3]{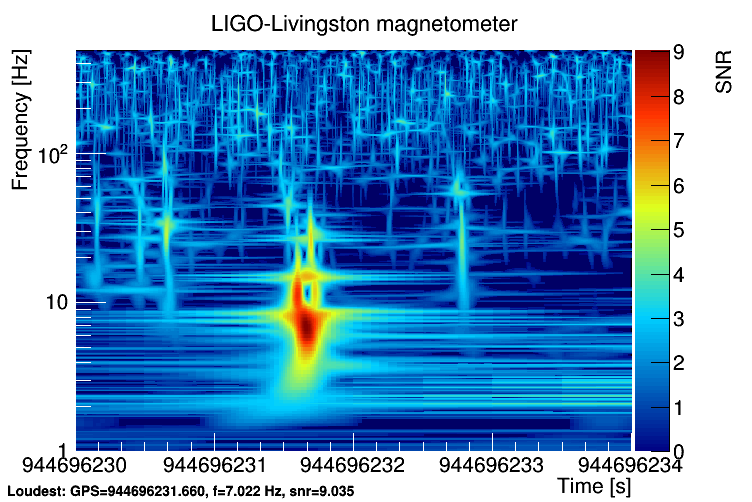}&
      \includegraphics[scale=0.3]{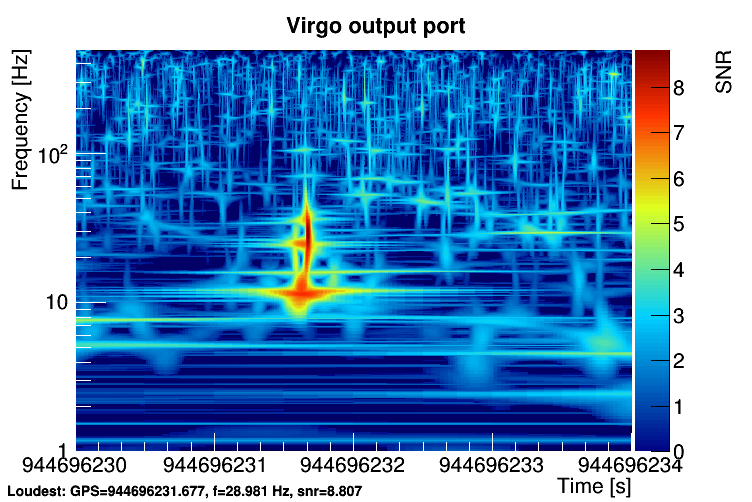}\\
    \end{tabular}
  \end{center}
  \caption{This figure presents the time-frequency spectrograms of magnetometers located at the LIGO-Hanford (Hanford, Washington, USA; magnetometer in the local $x$ direction; upper left), Virgo (Cascina, Italy; magnetometer in the local $y$ direction; upper right), and LIGO-Livingston (Livingston, Louisiana, USA; magnetometer in the local $z$ direction; bottom) sites at the time of the December 12, 2009 positive Gigantic Jet (at 23:36:56.55 UTC, or 944696231 GPS)~\cite{JGRD:JGRD16613}. The signal is clearly present in the LIGO-Hanford, LIGO-Livingston and Virgo magnetometers. In addition, the bottom right spectrogram shows the event in the Virgo gravitational-wave strain, h(t), data. Note that the SNR scales for the four figures are different, and the signal is strongest in the Virgo and LIGO-Hanford magnetometers. The observed events are at time consistent with other observations around the world~\cite{JGRD:JGRD16613}.}
  \label{fig:magnetometers}
\end{figure}

It is now clear that common magnetic field noise from the Schumann resonances~\cite{Sch1951,Sen1996} may very well inhibit the attempt by Advanced LIGO and Advanced Virgo to measure or set limits on the strength of a stochastic gravitational-wave background~\cite{2013PhRvD..87l3009T,PhysRevD.90.023013}
due to an undesired sensitivity of the instruments to magnetic fields. However, through the observations of the December 12, 2009 positive gigantic jet~\cite{JGRD:JGRD16613} it became apparent that LIGO and Virgo needed to worry about large magnetic transient events that could possibly create coincident short duration noise events at the different gravitational-wave detectors. The transient magnetic field from this event was simultaneously observed as a very loud signal in magnetometers located at the Virgo gravitational-wave detector, and the LIGO-Hanford observatory. A smaller signal was measured in a magnetometer at the LIGO-Livingston observatory. The time-frequency spectrograms of these magnetometer observations are displayed in Fig.~\ref{fig:magnetometers}.

What was particularly worrisome about this electromagnetic event was that it actually created a perceptible signal in Virgo's gravitational-wave strain (h(t)) channel; See Fig.~\ref{fig:magnetometers}, bottom right. No signal was definitively observed in the LIGO-Hanford and LIGO-Livingston h(t) data, although the spectrograms do show some low SNR structure that is coincident with the magnetic event.
It is problematic to know that a transient electromagnetic signal of global extent could produce a signal in a gravitational-wave detector.
Fortunately electromagnetic events of this magnitude will be detectable by the network of magnetometers at the gravitational-wave detector sites~\cite{0264-9381-33-13-134001}. 
In fact, the philosophy that Advanced LIGO~\cite{0264-9381-32-7-074001,0264-9381-27-8-084006} and Advanced Virgo~\cite{0264-9381-32-2-024001} have pursued for environmental monitoring is to place networks of detectors (magnetometers, seismometers, accelerometers, microphones, etc), more sensitive than the gravitational-wave detectors themselves to that particular noise source, for all environmental influences. As noted above, this Gigantic Jet event served as the motivation for the study presented in this paper. We are seeking to find out how often electromagnetic events can be observed in coincidence at global distances, and what are the amplitudes for these magnetic transients. It should be noted that the magnetic isolation for Advanced Virgo~\cite{0264-9381-32-2-024001} will probably be better than initial Virgo, and it would be unlikely for an event like this to be observed in Advanced Virgo. Advanced LIGO's~\cite{0264-9381-32-7-074001,0264-9381-27-8-084006} magnetic isolation has been measured to be of order \unit[$10^{-8}$]{m/T} from \unit[10 -- 20]{Hz} (and decreases to \unit[$10^{-9}$]{m/T} from \unit[25 -- 40]{Hz})~\cite{0264-9381-33-13-134001}. This implies that a \unit[100]{pT} (a large but not rare electromagnetic event~\cite{RDS:RDS5799}) magnetic pulse could move an Advanced LIGO mirror by approximately \unit[$10^{-18}$]{m}. For comparison, the gravitational wave observed by Advanced LIGO, GW150914, moved the LIGO mirrors by \unit[$2 \times 10^{-18}$]{m}~\cite{PhysRevLett.116.061102}. Hence it will be necessary to monitor coincident magnetic transient events, and especially to identify them and veto these times from the gravitational-wave searches.

\section{Analysis}
\label{sec:analysis}
To search for coincident short-duration magnetic events, we analyzed the magnetometer data using an algorithm called Omicron~\cite{Robinet:2009}. Derived from the original burst Q-pipeline~\cite{Chatterji:2004}, Omicron identifies and characterizes excess power noise transients in Advanced LIGO and Advanced Virgo detectors data~\cite{0264-9381-33-13-134001}. In addition, a gravitational-wave burst search is conducted using Omicron to detect transient signals in the output port of the LIGO and Virgo detectors~\cite{Lynch:2015yin,TheLIGOScientific:2016uux}. Omicron was designed to unite the robustness and noise parameter estimation accuracy of a burst search with the computational efficiency required for a near real-time analysis of several hundred channels on available computational resources. Omicron produces triggers using the Q transform~\cite{Brown:1991} which consists of projecting a data time series, $x(t)$, onto a basis of windowed complex exponentials defined by a central time $\tau$, a central frequency $\phi$, and a quality factor $Q$:
\begin{equation}
  X(\tau, \phi, Q) = \int_{-\infty}^{+\infty} x(t) w(t-\tau, \phi, Q) e^{-2i\pi\phi t} dt.
\end{equation}
The transform coefficient, $X$, measures the average signal amplitude and phase within a time-frequency region, called a tile, centered on time $\tau$ and frequency $\phi$, whose shape and area are determined by the requested quality factor $Q$ and the particular choice of analysis window, $w$. Although the optimal time and frequency resolution is achieved by a Gaussian window, for implementation purpose, a Connes window is implemented in Omicron. The tiles are distributed to cover a finite region in central time, central frequency, and $Q$ such that mismatch between any sinusoidal Gaussian in this signal space and the nearest basis function does not exceed a specified maximum mismatch. This naturally leads to a tiling structure consisting of logarithmically spaced $Q$ planes, logarithmically spaced frequency rows, and linearly spaced times. For each tile, a SNR value, $\rho$, is estimated. It measures the ratio between the signal and noise amplitude around time $\tau$ and frequency $\phi$. An Omicron trigger is defined as a tile the SNR of which is larger than a specified SNR threshold.

Omicron was configured to search events associated to the very specific phenomena that we were investigating. The frequencies of the Schumann resonances and ELF lightning-induced transients are located in the low end of the LIGO/Virgo sensitivity band, therefore we focused on low frequencies in our analysis, from 1 to 100 Hz. Moreover, the focus was set to the most significant events selected with $\rho > 5$. Multiple magnetometers are located around the LIGO and Virgo sites. For this study we analyzed the MFS-06 Metronix magnetometers at Virgo, the Bartington Mag-03 magnetometers at the LIGO-Hanford and LIGO-Livingston sites, and the custom built coil magnetometers at the LIGO-Hanford and LIGO-Livingston sites. The time period studied was during LIGO's sixth science run, and Virgo's second science run, from October 19, 2009, to January 8, 2010.

In order to find magnetic transient events observed at more than one gravitational-wave observatory simultaneously we conducted coincidence tests between magnetometer events seen at each pair of our detector locations.
The data are stamped using GPS and have a timing precision better than \unit[1]{$\mu$s}, far better than the accuracy required for this study.
Because of the broad structure, both in time and frequency, of the events caused by lighting discharges, we set a relatively wide time window for the coincidence test of \unit[0.25]{s}. We only studied triggers where the Omicron SNR exceeded 5 when the frequency range was restricted between \unit[7]{Hz} and \unit[25]{Hz}; 
aside from this restriction, the frequency content of the recorded events were not considered in determining coincidence. To estimate the background of random coincidences, we used standard procedure known as {\it time slides}. An artificial time delay 
greater than the propagation time of the signal
is inserted between the two data streams; in this way it is possible to know that all of the coincident events are due to noise.
We performed 80 time slides of the data, ranging between \unit[-5 and 5]{s}. 

\section{Results}
\label{sec:results}
\subsection{S6, VSR2 data}
We have analyzed all combination of different magnetometer channels from the three different detectors. Most of the combinations did not show any discernible effect. The total number of coincidences
was consistent with background. However, there were a few interesting combinations, where we have observed a significant excess of coincidences in zero lag (namely a true coincidence as opposed to what is seen with an artificial time delay) with
respect to other time slides. The presence or lack of an observable correlation depends on the level of the local electromagnetic fields, and the sensitivity of the magnetometers. 
In general the MFS-06 Metronix magnetometers at Virgo are more sensitive than the Bartington Mag-03 magnetometers at LIGO.

\begin{figure}
  \begin{center}
    \includegraphics[scale=0.75]{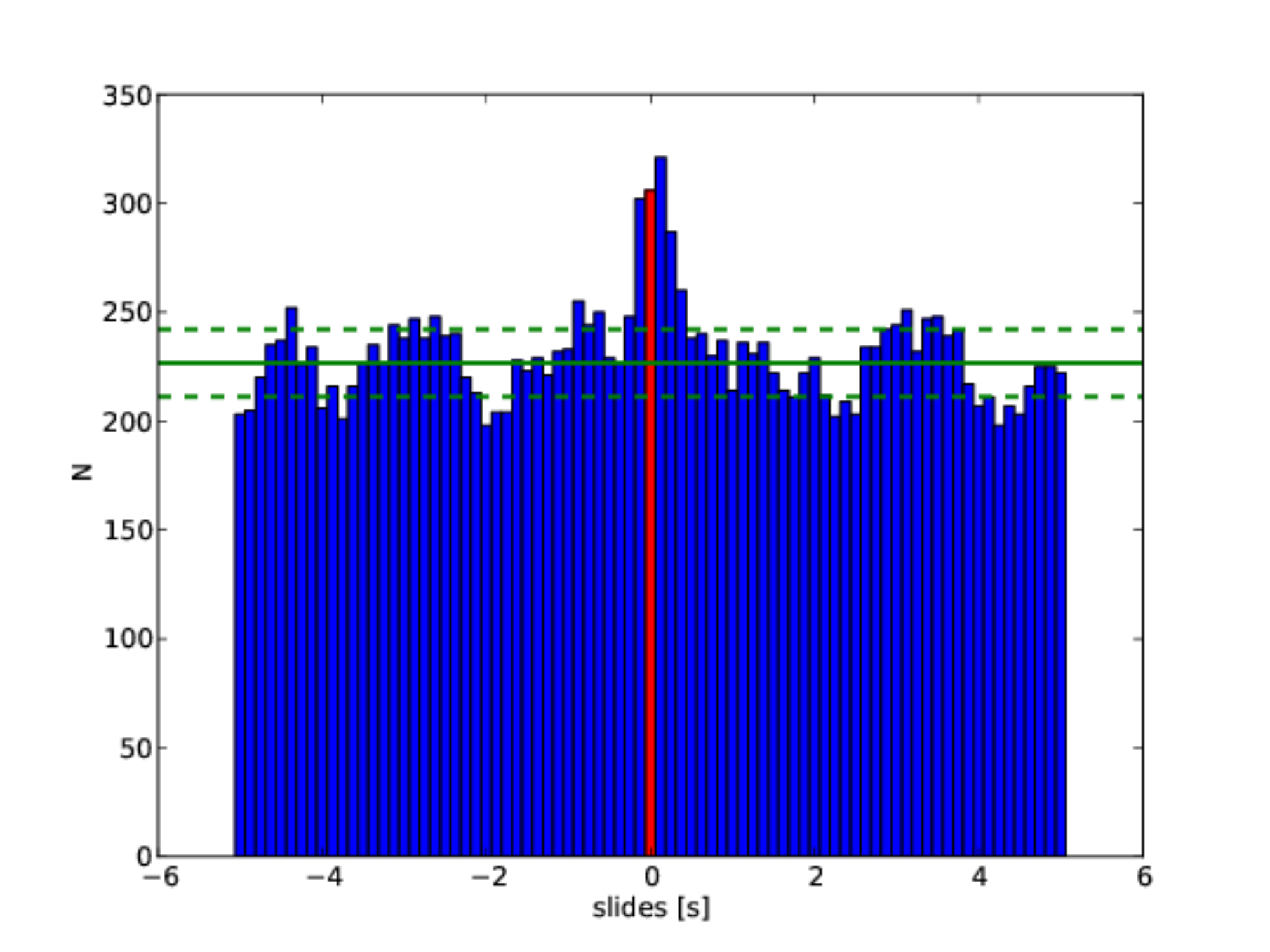}
  \end{center}
  \caption{The number of coincident triggers as a function of time delay for magnetometers located at LIGO-Hanford (magnetometer in the local $x$ direction) and LIGO-Livingston (magnetometer in the local $z$ direction). The horizontal line represents the mean value of the time slide results (excluding the \unit[0.625]{s} covered by the central 5 bins), while the dashed lines represent the standard deviation (again excluding the central 5 bins).}
  \label{fig:hist1}
\end{figure}

\begin{figure}
  \begin{center}
    \includegraphics[scale=0.75]{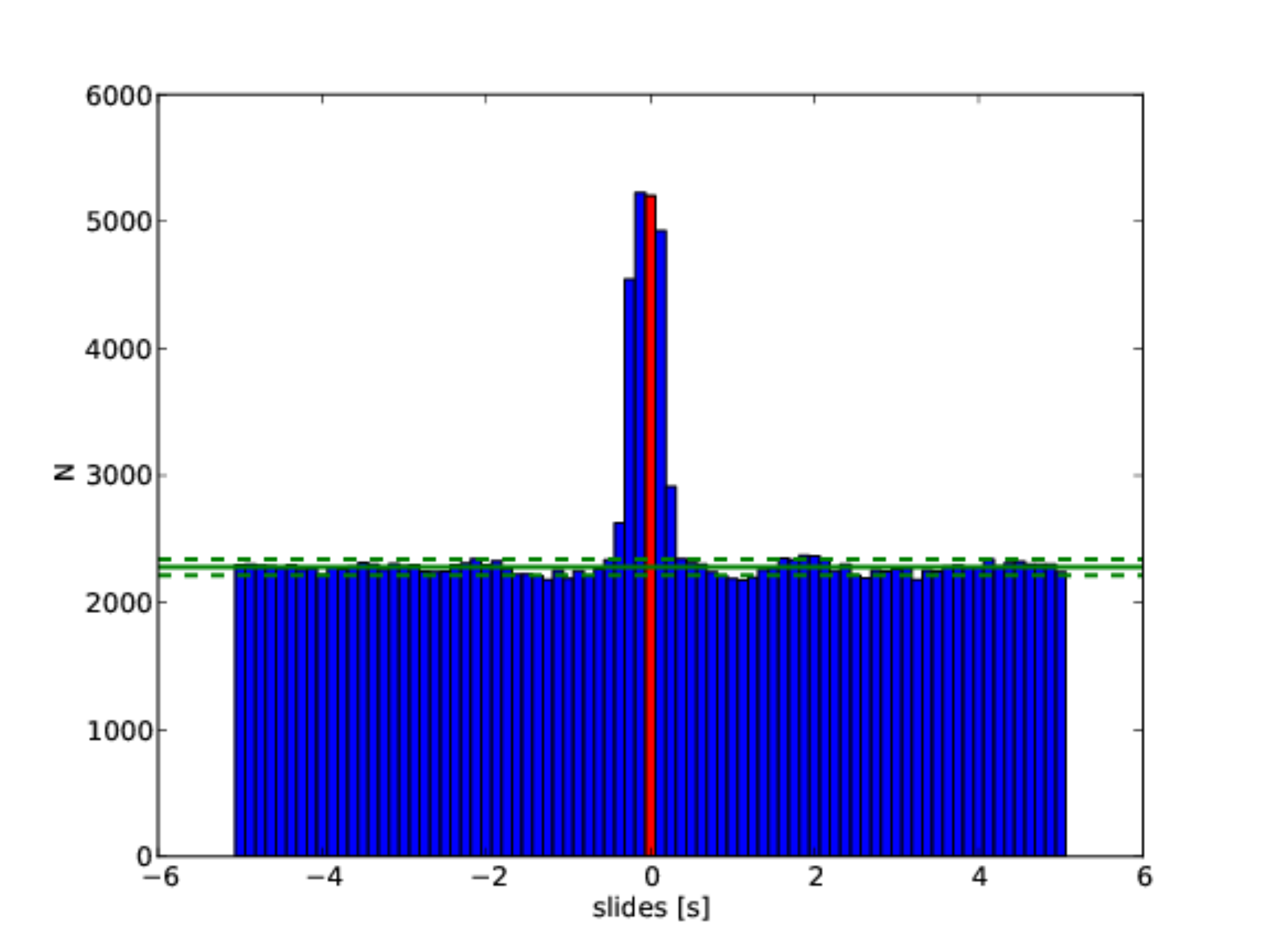}
  \end{center}
  \caption{The number of coincident triggers as a function of time delay for magnetometers located at LIGO-Hanford (magnetometer in the local $z$ direction) and Virgo (magnetometer in the local $x$ direction). The horizontal line represents the mean value of the time slide results (excluding the \unit[0.625]{s} covered by the central 5 bins), while the dashed lines represent the standard deviation (again excluding the central 5 bins).}
  \label{fig:hist2}
\end{figure}

Histograms showing the most interesting correlations are presented in Fig.~\ref{fig:hist1} for the LIGO-Hanford -- LIGO-Livingston combination, while Fig.~\ref{fig:hist2} displays a correlation between Virgo and LIGO-Hanford. Note that we were unable to measure any significant correlations between magnetometer data from Virgo and LIGO-Livingston. The distance between LIGO-Hanford and LIGO-Livingston is \unit[3030]{km} (\unit[10]{ms} light travel time), while the distance between LIGO-Hanford and Virgo is \unit[8850]{km} (\unit[30]{ms} light travel time). Both plots show an excess in the 5 central bins corresponding to a width of \unit[0.625]{s}.

\subsection{Poland-Colorado data}
The magnetic environments around the LIGO and Virgo sites are quite noisy. In order to better understand and measure correlated magnetic transient events on a global distance scale we analyzed data from very low noise magnetometers that have been installed in electromagnetic quiet locations.
Specifically, we used the ELF magnetometers from the Poland ($49.2^{\circ} N, 22.5^{\circ} E$) and Colorado ($38.9^{\circ} N, 103.4^{\circ} W$)
stations of the WERA project~\cite{WERA}. 
The distance between these stations is \unit[8873]{km} (\unit[30]{ms} light travel time). 

\begin{figure}
  \begin{center}
    \includegraphics[scale=0.75]{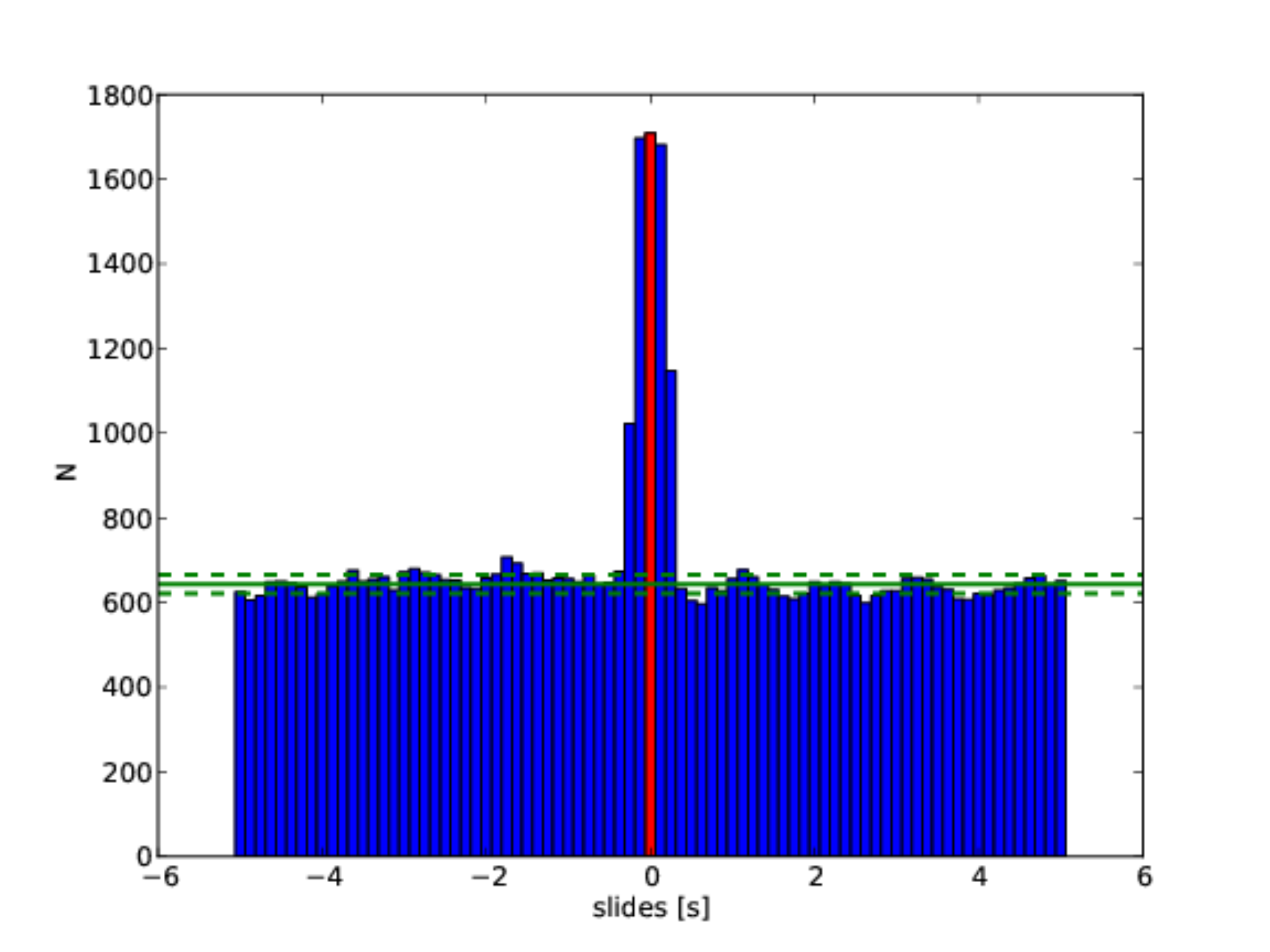}
  \end{center}
  \caption{The number of coincident triggers as a function of time delay for the low noise magnetometers located in Bieszczady Mountains in Poland (magnetometer in the local East - West direction) and the other in the Hugo State Wildlife Area in Colorado (magnetometer in the local East - West direction). The horizontal line represents the mean value of the time slide results (excluding the \unit[0.625]{s} covered by the central 5 bins), while the dashed lines represent the standard deviation (again excluding the central 5 bins).}
  \label{fig:hist3}
\end{figure}

We have analyzed data from the Poland--Colorado magnetometers over the period of June 3, 2015 to 
June 16, 2015 (13 days, 1 hour, 34 minutes, 32 seconds in total). Fig.~\ref{fig:hist3} displays the number of coincident triggers as a function of time lag. There is a clear excess of events at the zero lag time.

In order to better understand the potential deleterious effects that these coincident magnetic transient events might have on LIGO and Virgo data we made a histogram of the amplitude of the coincident events observed simultaneously in Poland and Colorado. The calibrated magnetometer data were passed through a \unit[1]{Hz} high-pass filter, and the peak (absolute value) value within $\pm$\unit[1]{s} of the Omicron trigger was selected. 

\begin{figure}
  \begin{center}
    \includegraphics[scale=0.70]{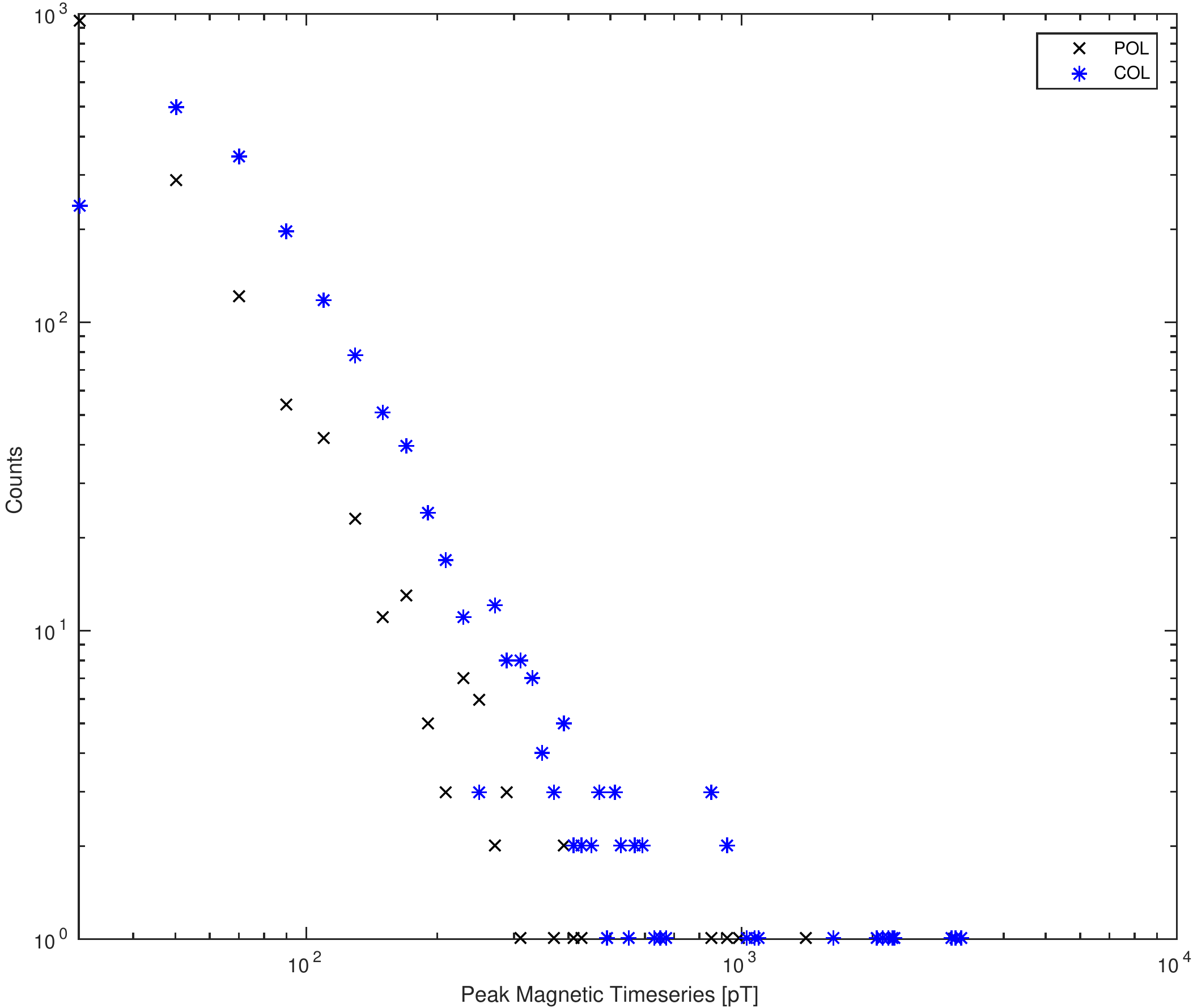}
  \end{center}
  \caption{A histogram of the amplitudes of the coincident magnetic triggers from the low noise magnetometers in Poland and Colorado (as displayed in Fig.~\ref{fig:hist3}. The calibrated magnetometer was high-pass filtered (\unit[1]{Hz}), and the peak (absolute value) value within $\pm$\unit[1]{s} of the Omicron trigger was selected.}
  \label{fig:hist4}
\end{figure}

Note that the shape of the distribution is similar for the magnetometers in Poland and Colorado, but the Colorado events are slightly larger. This could be due to a slight mis-calibration of the magnetometers, or proximity of the receiver to numerous lightning storms during the month of June. No attempt has been made to subtract off local magnetic noise from the pulses having a global extent. As such, the histograms in Fig.~\ref{fig:hist4} are likely to be a small overestimation of the globally coincident magnetic transient events.
Because of the tuning of the Omicron pipeline, there was a predisposition to identify magnetic events in the \unit[6--25]{Hz} band, where Advanced LIGO and Advanced Virgo will have good sensitivity and the presence of the Schumann resonances and ELF lightning-induced transients is the largest. This is also the region where the recently measured magnetic noise coupling~\cite{0264-9381-33-13-134001} is the largest, \unit[$10^{-8}$]{m/T} around \unit[10--20]{Hz}. We use these low noise measurement results from Poland and Colorado as an example of the globally coincident transient magnetic events that would effect LIGO and Virgo. A \unit[200]{pT} magnetic event could produce a signal in Advanced LIGO equivalent to a motion of the test mass of
\unit[$2 \times 10^{-18}$]{m}, the same amount that they moved when detecting GW150914~\cite{PhysRevLett.116.061102}. There were 30 events of this magnitude observed with the Poland magnetometer over the 13-day period, or roughly 2.3 per day. The rate was about twice as large in Colorado.

\subsection{Examination of coincident magnetic transients}
\begin{figure}
  \begin{center}
    \begin{tabular}{cc}
      \includegraphics[scale=0.4]{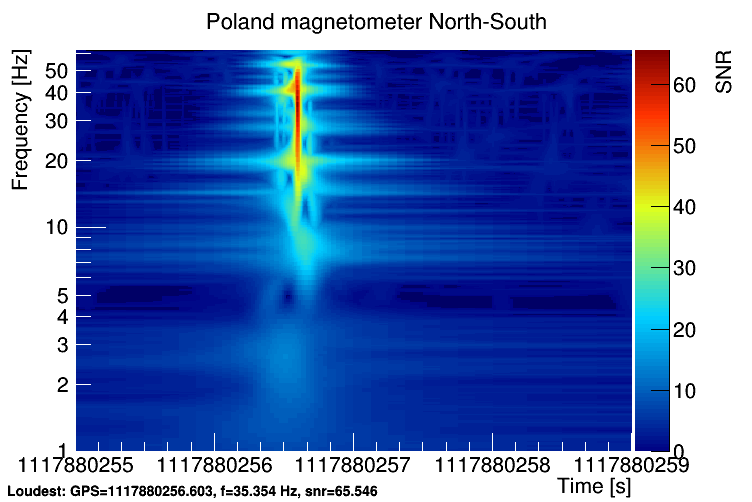} \\
      \includegraphics[scale=0.4]{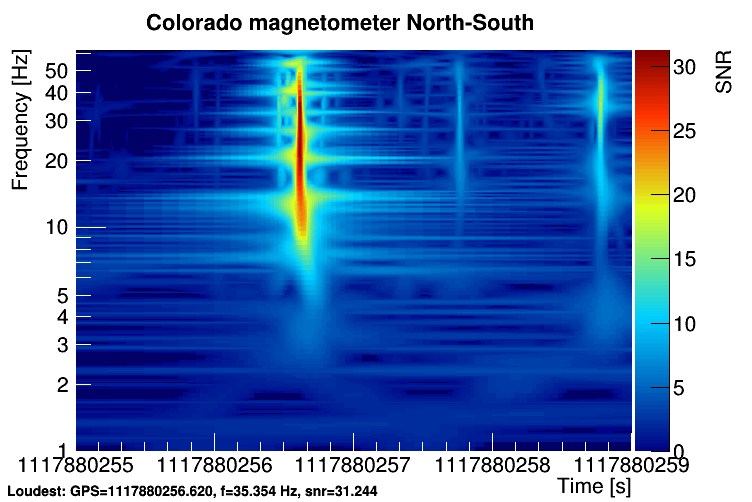} \\
    \end{tabular}
  \end{center}
  \caption{Displayed here is an example of a magnetic transient event that was observed simultaneously in the low-noise magnetometers in Poland and Colorado. The top figure displays the time-frequency spectrogram from the Poland North-South magnetometer data, while the bottom figure gives the time-frequency spectrogram for the Colorado North-South magnetometer data. Note the simultaneity and similarity of the signals. This event has been identified as occurring at 2015-06-09 at 10:17:20:563 UTC in China with latitude $= 23.942^{\circ}$, longitude $= 110.990^{\circ}$, with a peak current of \unit[149]{kA} and positive polarity.}
  \label{fig:example1}
\end{figure}

\begin{figure}
  \begin{center}
    \begin{tabular}{c}
      \includegraphics[scale=0.6]{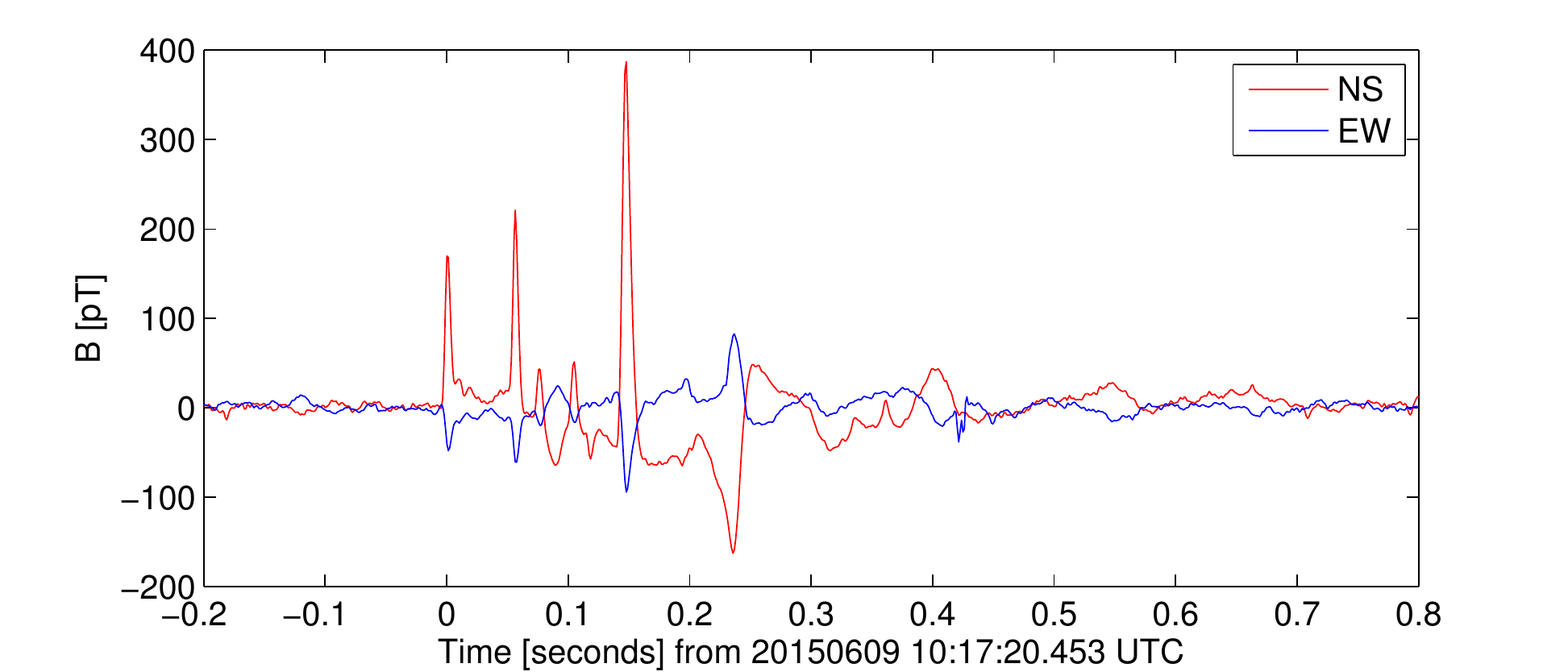}\\
      \includegraphics[scale=0.6]{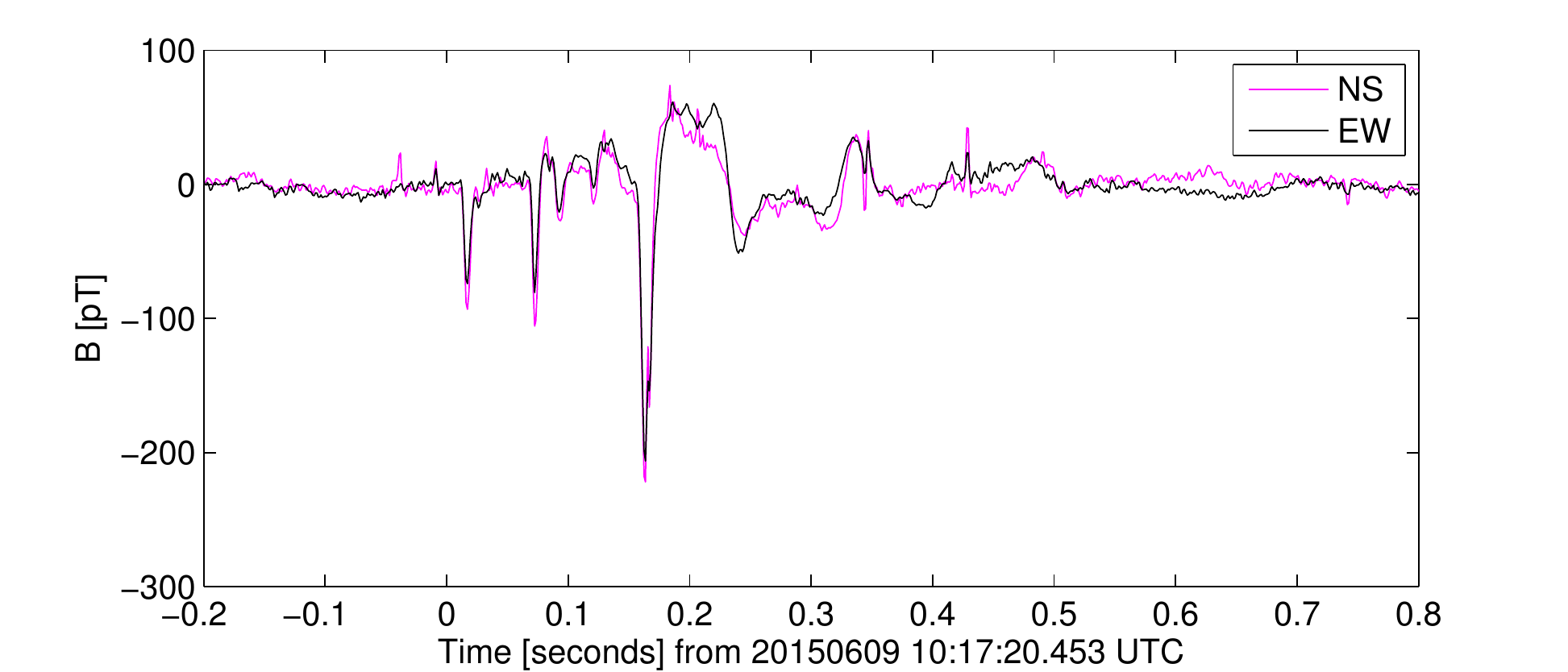}
    \end{tabular}
  \end{center}
  \caption{Displayed here the time series for the same magnetic event as in Fig.~\ref{fig:example1} that was observed simultaneously in the low-noise magnetometers in Poland and Colorado. The top figure displays the time series from the Polish North-South and East-West magnetometers, while the bottom plot is from the Colorado North-South and East-West magnetometers. A \unit[1]{Hz} high-pass filter has been applied to the data. The vertical axes are in units of pT. The strongest discharge has been identified as occurring at 2015-06-09 at 10:17:20:563 UTC in China with latitude $= 23.942^{\circ}$, longitude $= 110.990^{\circ}$, with a peak current of \unit[149]{kA} and positive polarity.}
  \label{fig:example1b}
\end{figure}

\begin{figure}
  \begin{center}
    \begin{tabular}{cc}
      \includegraphics[scale=0.4]{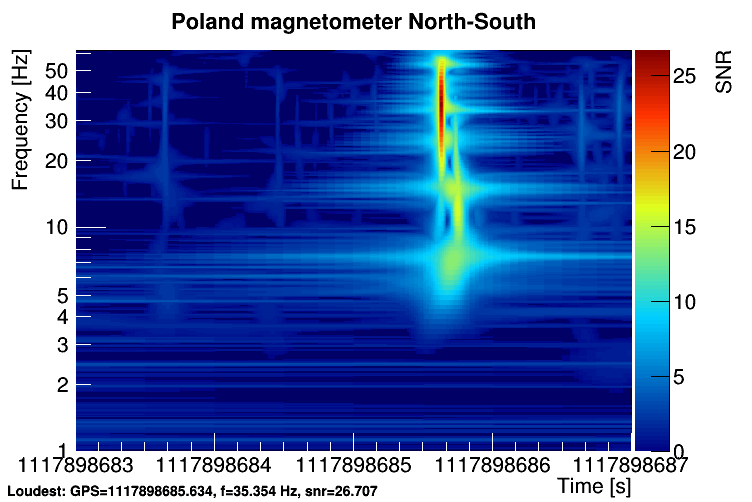} \\
      \includegraphics[scale=0.4]{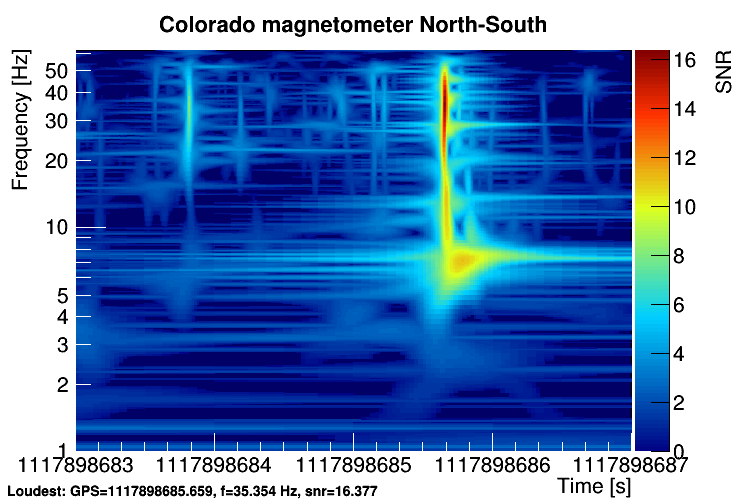} \\
    \end{tabular}
  \end{center}
  \caption{Displayed here is an example of a magnetic transient event that was observed simultaneously in the low-noise magnetometers in Poland and Colorado. The top figure displays the time-frequency spectrogram from the Poland North-South magnetometer data, while the bottom figure gives the time-frequency spectrogram for the Colorado North-South magnetometer data. Note the simultaneity and similarity of the signals. A \unit[1]{Hz} high-pass filter has been applied to the data. This event has been identified as occurring at 2015-06-09 at 15:24.29.599 UTC in India UTC with latitude $= 23.00^{\circ}$, longitude $= 88.42^{\circ}$.}
  \label{fig:example2}
\end{figure}

\begin{figure}
  \begin{center}
    \includegraphics[scale=0.6]{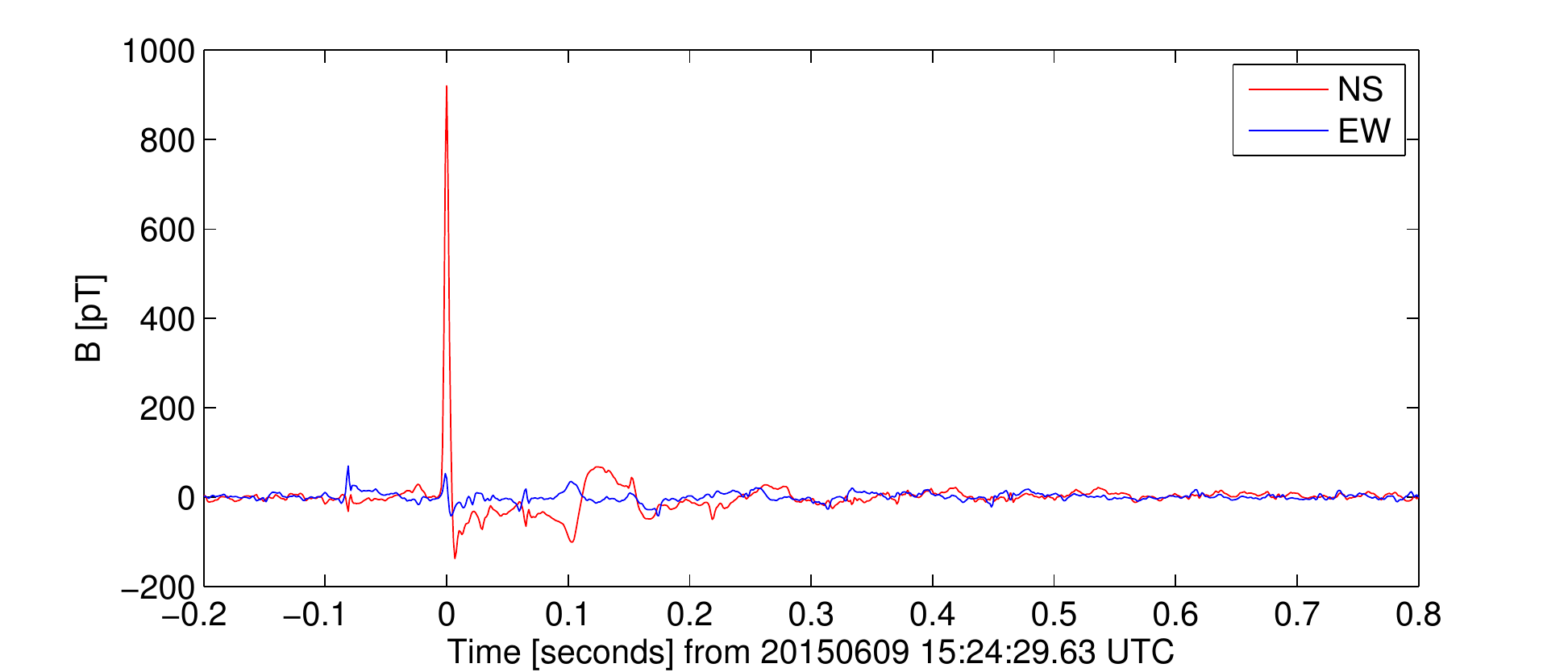}\\
    \includegraphics[scale=0.6]{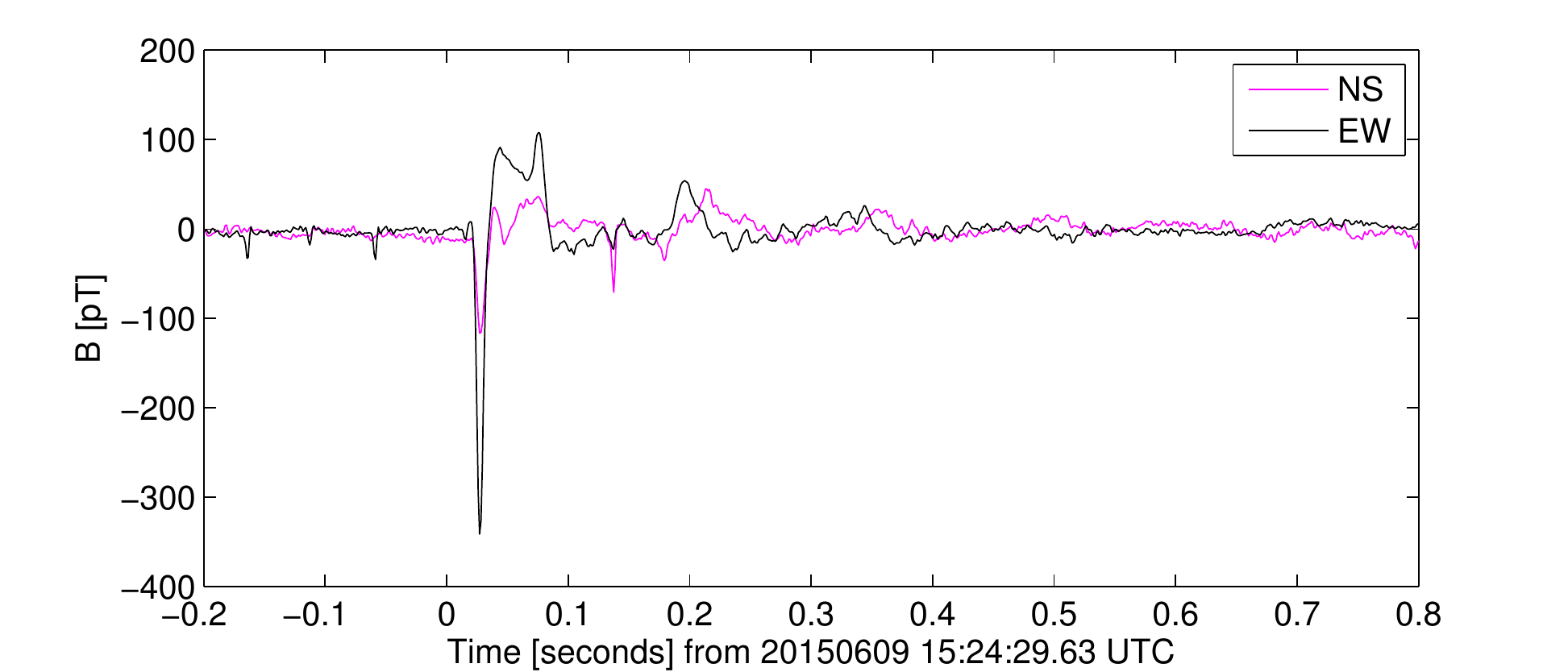}
  \end{center}
  \caption{Displayed here are the time series for the same magnetic event as in Fig.~\ref{fig:example2} that was observed simultaneously in the low-noise magnetometers in Poland and Colorado. The top figure displays the time series from the Poland North-South and East-West magnetometers, while the bottom plot is from the Colorado North-South and East-West magnetometers. A \unit[1]{Hz} high-pass filter has been applied to the data. The vertical axis is in unit of pT. This event has been identified as occurring at 2015-06-09 at 15:24.29.599 UTC in India UTC with latitude $= 23.00^{\circ}$, longitude $= 88.42^{\circ}$.}
  \label{fig:example2b}
\end{figure}

We examined a number of the magnetic transient events that were simultaneously observed in the Polish and Colorado magnetometers. Two examples are presented in Fig.~\ref{fig:example1} and Fig.~\ref{fig:example2}. The signals were observed in coincidence with the Polish and Colorado low-noise magnetometers. Like these signals, many of the observed coincident magnetic transients had amplitudes of hundreds of pT. Using the VAISALA GLD360 Global Lightning Dataset~\cite{JGRD:JGRD50508,JGRD:JGRD16376} and the WWLLN lightning data~\cite{Rodger-2009} we were able to identify the sources of these signals. The strongest of the impulses shown in Fig.~\ref{fig:example1} and Fig.~\ref{fig:example1b} has been identified as occurring at 2015-06-09 at 10:17:20:563 UTC in China with latitude $= 23.942^{\circ}$, longitude $= 110.990^{\circ}$, with a peak current of \unit[149]{kA} and positive polarity. 
The signal shown in Fig.~\ref{fig:example2} and Fig.~\ref{fig:example2b} occurred on the same day at 15:24.29.599 UTC in India (latitude $= 23.00^{\circ}$, longitude $= 88.42^{\circ}$).
Other coincidently detected events were also examined, and the source locations span the globe. These two events illustrate that it is possible to have globally coincident magnetic transient events with magnitudes in excess of hundreds of pT that happen multiple times each day.

\section{Conclusions}
\label{sec:conclusions}
One of the reasons why a global network of gravitational-wave detectors is needed is to be confident that there is not any common noise that could mimic a gravitational-wave signal. This has been the assumption used when analyzing LIGO and Virgo data in the past. However, it has been recently demonstrated that the magnetic fields associated with the Schumann resonances could corrupt the LIGO--Virgo effort to observe a stochastic background of gravitational waves~\cite{2013PhRvD..87l3009T,PhysRevD.90.023013}. What we have shown in the present paper is that magnetic transient events can be observed in coincidence on global distances. 
We have targeted magnetic transient events in the \unit[6--25]{Hz} band because this is where Advanced LIGO and Advanced Virgo will eventually have good sensitivity to gravitational waves, and it is also the region where the magnetic noise coupling to the interferometer test masses in the largest.
As such, these sorts of coincident noise events should merit careful monitoring for LIGO--Virgo for their short duration transient searches~\cite{PhysRevD.87.022002,PhysRevD.85.122007}.
Note that KAGRA's underground location does not shield it from magnetic noise from the Schumann resonances~\cite{1742-6596-716-1-012020}.

Going forward, Advanced LIGO and Advanced Virgo will have low noise magnetometers installed at each observatory site. It will be important to use the magnetometer data to identify coincident magnetic transient events, and veto those times. In fact, since these events are global in extent, it would also be prudent to use data from low noise magnetometers in low electromagnetic noise environments (such as the Poland and Colorado magnetometer used in this present study) to also identify globally coincident magnetic transients, and develop vetoes from those results. As the advanced detectors approach their design sensitivities it will be necessary to account for sources of correlated noise (short and long duration) between different sites.
Correlated noise will be a real and legitimate concern for gravitational-wave searches using ground based detectors.

\section{Acknowledgements}
NC and MR are supported by National Science Foundation (NSF) grant PHY-1505373 to Carleton College. 
MG acknowledges support from the NSF under Award AGS 1451210 to University of Colorado Denver. 
MC is supported by the NSF Graduate Research Fellowship Program, under NSF grant number DGE 1144152.
TB and IK are supported by NCN grant 2014/15/Z/ST9/00038. We thank D. Shoemaker, R. Schofield, M. Lintz and G. Bogaert for comments on the manuscript. This article has been assigned LIGO Document number P1600306.

\section*{References}
\bibliography{schumann}

\providecommand{\newblock}{}
\begin{thebibliography}{10}
\expandafter\ifx\csname url\endcsname\relax
  \def\url#1{{\tt #1}}\fi
\expandafter\ifx\csname urlprefix\endcsname\relax\def\urlprefix{URL }\fi
\providecommand{\eprint}[2][]{\url{#2}}

\bibitem{0264-9381-32-7-074001}
Aasi J {\em et~al.\/} (LIGO Scientific Collaboration and Virgo Collaboration)
  2015 {\em Classical and Quantum Gravity\/} {\bf 32} 074001

\bibitem{0264-9381-27-8-084006}
Harry G~M {\em et~al.\/} (LIGO Scientific Collaboration) 2010 {\em Classical
  and Quantum Gravity\/} {\bf 27} 084006

\bibitem{0264-9381-32-2-024001}
Acernese F {\em et~al.\/} (Virgo Collaboration) 2015 {\em Classical and Quantum
  Gravity\/} {\bf 32} 024001

\bibitem{0264-9381-29-12-124007}
Somiya K 2012 {\em Classical and Quantum Gravity\/} {\bf 29} 124007

\bibitem{doi:10.1142/S0218271813410101}
Unnikrishnan C~S 2013 {\em International Journal of Modern Physics D\/} {\bf
  22} 1341010 (\textit{Preprint}
  \eprint{http://www.worldscientific.com/doi/pdf/10.1142/S0218271813410101})

\bibitem{PhysRevLett.116.061102}
Abbott B~P {\em et~al.\/} (LIGO Scientific Collaboration and Virgo
  Collaboration) 2016 {\em Phys. Rev. Lett.\/} {\bf 116}(6) 061102

\bibitem{PhysRevLett.116.241103}
Abbott B~P {\em et~al.\/} (LIGO Scientific Collaboration and Virgo
  Collaboration) 2016 {\em Phys. Rev. Lett.\/} {\bf 116}(24) 241103

\bibitem{PhysRevX.6.041015}
Abbott B~P {\em et~al.\/} (LIGO Scientific Collaboration and Virgo
  Collaboration) 2016 {\em Phys. Rev. X\/} {\bf 6}(4) 041015

\bibitem{0264-9381-27-19-194010}
Christensen N {\em et~al.\/} (LIGO Scientific Collaboration and Virgo
  Collaboration) 2010 {\em Classical and Quantum Gravity\/} {\bf 27} 194010

\bibitem{0264-9381-32-11-115012}
Aasi J {\em et~al.\/} (LIGO Scientific Collaboration and Virgo Collaboration)
  2015 {\em Classical and Quantum Gravity\/} {\bf 32} 115012

\bibitem{0264-9381-29-15-155002}
Aasi J {\em et~al.\/} (LIGO Scientific Collaboration and Virgo Collaboration)
  2012 {\em Classical and Quantum Gravity\/} {\bf 29} 155002

\bibitem{0264-9381-33-13-134001}
Abbott B~P {\em et~al.\/} (LIGO Scientific Collaboration and Virgo
  Collaboration) 2016 {\em Classical and Quantum Gravity\/} {\bf 33} 134001

\bibitem{2013PhRvD..87l3009T}
{Thrane} E, {Christensen} N and {Schofield} R~M~S 2013 {\em Phys. Rev. D\/}
  {\bf 87} 123009 (\textit{Preprint} \eprint{1303.2613})

\bibitem{PhysRevD.90.023013}
Thrane E, Christensen N, Schofield R~M~S and Effler A 2014 {\em Phys. Rev. D\/}
  {\bf 90}(2) 023013

\bibitem{PhysRevLett.113.231101}
Aasi J {\em et~al.\/} (LIGO Scientific Collaboration and Virgo Collaboration)
  2014 {\em Phys. Rev. Lett.\/} {\bf 113}(23) 231101

\bibitem{PhysRevD.46.5250}
Christensen N 1992 {\em Phys. Rev. D\/} {\bf 46}(12) 5250--5266

\bibitem{PhysRevD.87.022002}
Aasi J {\em et~al.\/} (LIGO Scientific Collaboration and Virgo Collaboration)
  2013 {\em Phys. Rev. D\/} {\bf 87}(2) 022002

\bibitem{PhysRevD.85.122007}
Abadie J {\em et~al.\/} (LIGO Scientific Collaboration and Virgo Collaboration)
  2012 {\em Phys. Rev. D\/} {\bf 85}(12) 122007

\bibitem{PhysRevD.93.042005}
Abbott B~P {\em et~al.\/} (LIGO Scientific Collaboration and Virgo
  Collaboration) 2016 {\em Phys. Rev. D\/} {\bf 93}(4) 042005

\bibitem{JGRD:JGRD16613}
van~der Velde O~A, B{\'o}r J, Li J, Cummer S~A, Arnone E, Zanotti F,
  F{\"u}llekrug M, Haldoupis C, NaitAmor S and Farges T 2010 {\em Journal of
  Geophysical Research: Atmospheres\/} {\bf 115} D24301 ISSN 2156-2202 d24301

\bibitem{6353166}
Kulak A and Mlynarczyk J 2013 {\em IEEE Transactions on Antennas and
  Propagation\/} {\bf 61} 2269--2275 ISSN 0018-926X

\bibitem{rakov2003lightning}
Rakov V and Uman M 2003 {\em Lightning: Physics and Effects\/} (Cambridge
  University Press) ISBN 9780521583275
  \urlprefix\url{https://books.google.fr/books?id=NviMsvVOHJ4C}

\bibitem{Sch1951}
Schumann W 1952 {\em Zeitschrift f{\"u}r Naturforschung A\/} {\bf 7} 149--154

\bibitem{balser1960observations}
Balser M and Wagner C 1960 {\em Nature\/} {\bf 188} 638--641

\bibitem{JGRA:JGRA51647}
Mlynarczyk J, Bor J, Kulak A, Popek M and Kubisz J 2015 {\em Journal of
  Geophysical Research: Space Physics\/} {\bf 120} 2241--2254 ISSN 2169-9402
  2014JA020780

\bibitem{RDS:RDS5799}
Kulak A and Mlynarczyk J 2011 {\em Radio Science\/} {\bf 46} n/a--n/a ISSN
  1944-799X rS2016

\bibitem{kulak2014extremely}
Kulak A, Kubisz J, Klucjasz S, Michalec A, Mlynarczyk J, Nieckarz Z, Ostrowski
  M and Zieba S 2014 {\em Radio Science\/} {\bf 49} 361--370

\bibitem{WERA}
 2016 {\em WERA Project\/} \\ \url{
  http://www.oa.uj.edu.pl/elf/index/projects3.htm}

\bibitem{JGRD:JGRD16034}
Kulak A, Nieckarz Z and Zieba S 2010 {\em Journal of Geophysical Research:
  Atmospheres\/} {\bf 115} n/a--n/a ISSN 2156-2202 d19104

\bibitem{pasko2010recent}
Pasko V~P 2010 {\em Journal of Geophysical Research: Space Physics\/} {\bf 115}

\bibitem{GRL:GRL28741}
Cotts B~R~T, Gołkowski M and Moore R~C 2011 {\em Geophysical Research
  Letters\/} {\bf 38} n/a--n/a ISSN 1944-8007 l24805

\bibitem{JGRD:JGRD18081}
Kulak A, Mlynarczyk J, Ostrowski M, Kubisz J and Michalec A 2012 {\em Journal
  of Geophysical Research: Atmospheres\/} {\bf 117} n/a--n/a ISSN 2156-2202
  d18203

\bibitem{PhysRevD.85.082002}
Abadie J {\em et~al.\/} (LIGO Scientific Collaboration and Virgo Collaboration)
  2012 {\em Phys. Rev. D\/} {\bf 85}(8) 082002

\bibitem{PhysRevD.89.122003}
Aasi J {\em et~al.\/} ((LIGO Scientific Collaboration and Virgo Collaboration))
  2014 {\em Phys. Rev. D\/} {\bf 89}(12) 122003

\bibitem{0004-637X-785-2-119}
Aasi J {\em et~al.\/} (LIGO Scientific Collaboration and Virgo Collaboration)
  2014 {\em The Astrophysical Journal\/} {\bf 785} 119

\bibitem{0264-9381-32-3-035017}
Effler A, Schofield R~M~S, Frolov V~V, Gonz{\'a}lez G, Kawabe K, Smith J~R,
  Birch J and McCarthy R 2015 {\em Classical and Quantum Gravity\/} {\bf 32}
  035017

\bibitem{Sen1996}
Sentman D~D 1996 {\em J. Geophys. Res.\/} {\bf 101} 9479--9487

\bibitem{Robinet:2009}
Robinet F 2016 {\em Omicron: an algorithm to detect and characterize transient
  events in gravitational-wave detectors\/} \\ \url{
  https://tds.ego-gw.it/ql/?c=10651}

\bibitem{Chatterji:2004}
Chatterji S 2004 {\em The search for gravitational wave bursts in data from the
  second LIGO science run\/} Ph.D. thesis MIT

\bibitem{Lynch:2015yin}
Lynch R, Vitale S, Essick R, Katsavounidis E and Robinet F 2015
  (\textit{Preprint} \eprint{1511.05955})

\bibitem{TheLIGOScientific:2016uux}
Abbott B~P {\em et~al.\/} (LIGO Scientific Collaboration and Virgo
  Collaboration) 2016 {\em Phys. Rev.\/} {\bf D93} 122004 (\textit{Preprint}
  \eprint{1602.03843})

\bibitem{Brown:1991}
Brown J~C 1991 {\em The Journal of the Acoustical Society of America\/} {\bf
  89} 425--434

\bibitem{JGRD:JGRD50508}
Said R~K, Cohen M~B and Inan U~S 2013 {\em Journal of Geophysical Research:
  Atmospheres\/} {\bf 118} 6905--6915 ISSN 2169-8996

\bibitem{JGRD:JGRD16376}
Said R~K, Inan U~S and Cummins K~L 2010 {\em Journal of Geophysical Research:
  Atmospheres\/} {\bf 115} n/a--n/a ISSN 2156-2202 d23108

\bibitem{Rodger-2009}
Rodger C~J, Brundell J~B, Holzworth R~H and Lay E~H 2009 {\em AIP Conference
  Proceedings\/} {\bf 1118} 15--20

\bibitem{1742-6596-716-1-012020}
Atsuta S, Ogawa T, Yamaguchi S, Hayama K, Araya A, Kanda N, Miyakawa O, Miyoki
  S, Nishizawa A, Ono K, Saito Y, Somiya K, Uchiyama T, Uyeshima M and Yano K
  2016 {\em Journal of Physics: Conference Series\/} {\bf 716} 012020

\end{thebibliography}
\end{document}